\documentclass[10pt,a4paper]{article}
\usepackage{amsmath,amssymb,amsthm}
\usepackage{array}
\usepackage{authblk}
\usepackage{booktabs}
\usepackage{caption}
\usepackage{enumitem}
\usepackage{float}
\usepackage{geometry}
\usepackage{graphicx}
\usepackage[utf8]{inputenc}
\usepackage{longtable}  
\usepackage{lmodern}
\usepackage{multirow}
\usepackage{setspace}
\usepackage{titlesec}
\usepackage[T1]{fontenc}
\usepackage{xcolor}
\usepackage{hyperref}
\usepackage{pdfpages} 
\titleformat{\section}{\normalfont\Large\bfseries}{\thesection.}{0.5em}{}
\titleformat{\subsection}{\normalfont\large\bfseries}{\thesubsection.}{0.5em}{}

\title{A3T-GCN for FTSE100 Components Price Forecasting}
\author{Andres Luis Paredes Ramos\textsuperscript{*}}
\date {}
\affil{\textsuperscript{*}Leeds University Business School, Leeds, UK}

\begin{document}
\maketitle

\begin{abstract} 
\noindent We examine the predictive power of a novel hybrid A3T-GCN architecture to forecast closing stock prices within constituents of the FTSE100 index. The final dataset included 79 companies and 375,329 observations, with node features including technical indicators (RSI, MACD), transformed returns (Normalized Returns, Log Returns) and annualized log returns over different windows (ALR1W, ALR2W, ALR1M, ALR2M). Graph composition relied on sector classifications with Pearson correlation of returns or Spearman correlation of financial ratios. We utilize UK stock market data via a proxy (FTSE100 index) covering daily periods from 2007-2024. This works finds that the A3T-GCN model using annualized log-returns as features and shorter sequence lengths improves accuracy while decreasing the computing power required. Additionally, it shows that longer historical contexts result in smaller increases in prediction errors, indicating their necessity when making more distant future predictions. 
\end{abstract}

\pagebreak

\section{Introduction}
Confidence in accurate forecasted returns are the catalyst behind the execution of millions of bid-ask operations in
the open market that reached a total stock traded value of 104.29 trillion USD in 2022 alone (World Bank, 2024).
Many factors influence the security selection decision making, with Poterba’s (2001) six portfolio behaviours,
driven by tax considerations; alongside Markowitz (1952) Modern Portfolio Theory, emphasizing risk tolerance as
critical in portfolio building, being foundational works for poising investors assessments. Nevertheless, rational
profit-maximising investors, encompassed in asset managers, retail or institutional traders, rather compete by
choosing to buy or sell a security based on their forecast that a company’s value will be worth more or less
tomorrow (Fama, 1970). The inherent pressure of instant, accurate forecast making, combined with the vast amount
of information to be considered, meant that across the market cycles quantitative methods have become preferred for
their capability of handling an instant analysis of data (Yoshihara, 2015).\par
While an increase in computational power facilitates faster analysis, McNelis (2005) highlights that "the specific
way of conceptualizing problems continues to play an important role in how quickly reliable results may be
obtained." This view fosters our aim on studying a novel untested hybrid A3T-GCN architecture for the prediction of
closing stock prices. We address the complexities of traditional financial time series prediction; in particular, the
high degree of noise (Brogaard, 2022), non-linear factors (Liu, 2024; Amini et al., 2021) and the established
convention of semi-strong market efficiency hypothesis (Fama, 1970). Moreover, elucidating the longstanding
difficulty in outperforming the market, classic work by Jensen (1968) shows, after transaction costs and at 2 t-
statistics, only 1 of 115 mutual funds presented abnormal “alphas” during 1945-1964. More recently the SPIVA US
mid-year report of 2024 showed that 57\% active large-cap equity managers underperformed the S\&P 500 (Standard \& Poor's, 2024).\par
GNNs have emerged as a promising approach in neural networks for graph-based task involving relational data.
Recent literature showed various success in various applications. Such as, forecasting of solar power generation, sign language translation, epidemic forecasting or traffic speed prediction (Simeunović et al., 2021; Kan et al., 2021; Wang et al., 2022; Shen et al., 2022). Despite rapid development inside the field, stock price prediction remained a nascent area of research. In this respect, we make two contributions to the literature: \\
First, we focus on an attentional spatial network (A3T-GCN) with propagation module via a convolution operator. Interestingly, there has been no prior effort to implement this hybrid model on a large and liquid stock universe, such as the FTSE100, to evaluate its performance in financial prediction tasks. We fill this void and apply our model on daily data obtained from Yahoo Finance, covering the period of 2007 until 2024. We partition on a 90:10 basis for training and testing sets.\\
Second, we aim at discussing the theoretical constraint of variable explainability in relation to the source of predictions. We find that the field of study still lacks explainability methodologies similar to Integrated Gradients
(Sundararajan et al., 2017) used for CNNs. Dai et al. (2023) points out that despite advancements in explainable DNNs, their focus has been on images and text, remaining unapplicable to GNNs due to the message-passing of GNNs and the discreteness of graph topology. \par
Our key findings show that using variants of ALR features (annualized log-returns across different time frames) reduces computing power by replacing the need for longer sequence length while achieving comparable or better error metrics. Under those conditions, a shorter sequence length provides more accurate results, as they introduce
less noise in the model input data; nevertheless, a longer sequence length is necessary to prevent the degradation of predictive power when increasing the forecast horizon to more distant points in time. Finally, we find that the methodology by Yin et al. (2021) and Sun et al. (2020) offer better results than our proposed spearman fundamental ratios during the graph composition. \par
The remainder of this paper is organized as follows. Section 2 reviews the existing literature. Section 3 details the
methodology. Section 4 discusses the results. Section 5 concludes and presents our ethical considerations.
\section{Literature Review}
\subsection{Security selection performance}
Investing in the stock market has historically been and continues to be an activity used to generate wealth, as evidenced by the significant role of this asset class in portfolios. To illustrate, in the United States, over 60\% of wealth creation during the 1990s was attributed to rising household stock holdings (Poterba, 2000). Similarly, during 1963, a substantial portion of the London Stock Exchange's equity ownership was attributed to individuals (54\%) reaching 93\% when aggregated with other local financial institutions like insurers and pensions (Office for National
Statistics, 2022). Currently, from the theoretical USD 175.0 trillion in global investable assets, equities had a market capitalization of 44.8\% or USD 78.0 trillion (State Street, 2024); displaying the enduring importance of this security type and that of adequate active management investing. On the last point, Firth (1977) explains that if scare capital is to be used more efficiently, accurate pricing of securities is essential; unsurprisingly, stock-picking abilities remain highly coveted. \par
Interestingly, literature points towards the difficulty for most investors in beating the passive benchmark1. Under the EMH proposed by Fama (1970), in a risk-adjusted basis, investors performance should be random unless they are in possession of private information not yet known to the wider market. Indeed, despite an equal expected chance of over- or underperformance, most investors rather fail to beat the market by varying degrees. \par
For the case of mutual funds, the evidence overwhelmingly suggests that, in the past decades, most funds simply fail to outperform their passive benchmarks. Carhart (1997) landmark work supported the inexistence of skilled or informed mutual fund portfolio manager that would allow for a generation of superior returns. This is consistent with earlier work by Malkiel (1995), who attributed the underperformance of mutual funds from 1971 to 1999 to high expense fees and inefficient market timing strategies by managers. Chan et al. (2002) further reinforced these findings, demonstrating that most growth mutual funds closely track the market index, with only a minor outperformance of growth relative to value style funds. Gruber (1996) posits an explanation for the unceasing growth of the industry, finding that mutual funds are priced at their net asset values, without any changes in their quoted prices to reflect the skill or quality of management. This inefficiency in pricing allows underperforming funds to continue increasing their AUM, despite their inability to beat the low-cost index alternatives. Alternatively, Pastor and Stambaugh (2012) found that investors tend to believe in decreasing return to scale; meaning they support the view that size increases the difficulty in generating returns, adjusting their expectations as such towards tolerating more mediocre performance. Interestingly, later work by Pastor et al. (2014) provided a more technical view on size of funds, finding that for every 1\% increase in the size of the industry, mutual fund performance declined by 40 basis points; suggesting a wrongful initial assessment by investors. Unfortunately for investors, underperformance is also present during periods of downturns. Pastor and Blair’s (2020) analysis of mutual fund performance during the first 10 weeks of the COVID-19 crisis found that 74.2\% of mutual funds underperformed the S\&P 500, Fink et al. (2014) evidences the industry had already disappointed during earlier recessions, the 2008 financial crisis, the 2000 dot-com bubble, banking crises in early 1990s and during the oil price shock in the 1980s.
\par
Retail investors do not perform any better in comparison. Cremers et al. (2018) point to high search costs, behavioural biases, and limited access to information as the cause for an approximate 1 to 2\% underperformance relative to the market. Coval et al. (2018) finds a wider array of reasons for the systematic underperformance of 8%
annually for the bottom decile of household investors; including, misperception of market conditions, liquidity trading due to budget needs, high transaction costs, lack of sophistication in exploiting market inefficiencies, herding behaviour, overconfidence, and irrational decision making based on emotions. Further, Barber and Odean (2000) trace transactions costs as the main driver behind a 1.1\% annualized underperformance on their 66,456 household sample. They find overconfidence as the catalyst for the elevated number of trades, a condition further exacerbated by the disposition effect identified by Shefrin and Statman (1985), by which investors not only trade too frequently but also hold onto losing positions, worsening overall performance.

\subsection{Quantitative methods on price prediction}
In this context of financial time series forecasting tools, three key areas have been most extensively explored in the literature; LSTMs, CNNs and ARIMA. Studies have consistently shown an outperformance of LSTM over ARIMA; finding a significant improvement of 6 to 7 times more accuracy when using the former (Siami-Namini et al., 2018; Selvin et al., 2017). Yet, the hybrid models that combine CNN and LSTM achieve the highest accuracy and robustness when compared to single-model architectures (Vidal and Kristjanpoller, 2020; Lu et al., 2020) by leveraging CNN’s capacity for feature extraction together with LSTM’s strength in temporal memory. This raises a broader question about the effectiveness of GNNs in this context, which have already emerged as a promising approach in neural networks for graph-based task involving relational data, including forecasting solar power generation, sign language translation, epidemic estimation or traffic speed prediction (Simeunović et al., 2021; Kan et al., 2021; Wang et al., 2022; Shen et al., 2022). Nevertheless, price prediction remains a nascent area of research; with a limited understanding of how to best represent inter-stock returns relationships in the graph structure or which
features are most relevant for improving predictive accuracy. \par
From the many available architectures inside the field, see Appendix 1, this paper makes use of an A3T-GCN, an attention temporal GCN which for its characteristics we classify inside the spatial subcategory of GNNs. The original design for traffic forecasting was proposed by Zhu et al. (2020). This specific cell, see Appendix 2, was
chosen from the wide array of possibilities to simulate the findings of Vidal and Kristjanpoller (2020) and Lu et al. (2020) by replicating the feature extraction capacity of CNNs through a GCN component, while also including a GRU, a streamlined LSTM which operates on fewer parameters and outperforms in convergence speed and
generalization (Chung et al., 2014). Effectively imitating the CNN-LSTM hybrid in the field of graph neural networks.

\subsection{Model configuration}
With a chosen model, the question now resides in what to translate from real-world conditions into data for testing. Firstly, the selection of variables was informed by previous studies in the field. Sun et al., (2020) successfully utilized normalized daily return data in their MCT-GCN model; this feature serves as the conceptual anchor from which other related indicators are derived. Campbell et al. (1997) had highlighted the better statistical properties of log-returns, such as tendency for normality, stationarity and time additivity. The inclusion of both would allow the
model to capture different dimensions of price dynamics, normalized returns allow a relative cross asset understanding of price changes, while log returns capture the absolute magnitude of price change. The time additivity property is important for our purposes; given our computing power constraint, we posit if adding a feature
with long-term context (ALR) could reduce the need for a larger sequence length; at a minimum, it should offer the model to differentiate between short-term fluctuations and long-term trends. Finally, the results of Lee et al. (2021) highlighted RSI and MACD to have higher predictive power compared to other well-known technical indicators,
under a standalone LSTM for the TWSE 0050 index. Lo et al. (2000) had already concluded that technical indicators provide incremental information about stock returns, suggesting they can enhance predictive models; thus, these momentum indicators are included in our features too. \par
Sector classification is the most immediate proxy for a methodically examined relation amongst otherwise different companies. Further, it is widely available in most markets although not explicitly comparable unless based on similar standards. We make use of our market’s LSEG classification, the Industry Classification Benchmark, a system of 11 Industries; 20 Supersectors; 45 Sectors and 173 Subsectors. McNamara et al. (2005) demonstrated that sector classification explains approximately 9.1\% of the total variance in US firms’ performance, thereby highlining why we select this as our first component for the graph composition alongside the Pearson correlation of returns
based on previous work by Pillay and Moodley (2022), Yin et al. (2021), and Sun et al. (2020) on similar architectures based on GNN. A different proposition stems from the results of Lewellen (2004), as they demonstrate the strong predictive power of certain fundamental ratios for market returns. Given their common use among investors, as described in various sources (Nissim and Penman, 2001; Lipson, 2019; Lev and Sunder, 1979), these results motived us to explore a further graph composition based on this line of though. However, given the particular nature of this type of data we decided to use the Spearman correlation to address scale differences. Literature by Xin
et al. (2024) proved the capacity of spearman correlation but for quantifying similarity of historical returns trends in a StockGCN; Marti et al. (2021) further advocated for its use to filter out distributional effects. Consequently, our
alternative graph composition includes the spearman correlation of a random selection of 28 fundamental ratios from the directory of 138 available in Yahoo Finance. \par
Finally, for our sequence length, we follow the findings of French (1980) who first described the Friday or Weekend effect, which was later further investigated by Keim and Stambaugh (1984), Lakonishok and Maberly (1990), and Brusa et al. (2000). Empirical evidence supports the notion that investors tend to sell more on Mondays, we translate it as a necessity to include a full week (Monday-Friday or 5 days) of contextual information for the model prediction. A comparable effect is seen in the findings of Rozeff and Kinney (1976) for the January effect in relation
to tax-loss selling in December and a posterior reinvestment in January, and Ariel (1987) contending abnormal high returns at the end and beginning of a month. Therefore, we choose to explore this too by further testing a full month or 30-sequence length configuration.

\section{Data \& Methodology}
Three steps are required; obtaining our population, preprocessing the data in terms of a graph composition, and the training of learnable parameters across the designated hyperparameters.
\subsection{Construction of the dataset}
Three different datasets are required; concisely, the sector classification, the associated historical daily features, and fundamental key ratios. The study period is 01-01-2007 to 30-12-2024. Under a 90/10 train data proportion, therefore, from 01-01-2007 to 06-06-2023 (16 years) for training and 07-06-2023 to 30-12-2024 (1.5 years) for testing.
Procuring the sector classification is the initial step. Dataset 1 (sector classification) is obtained by data scraping the 'FTSE 100 Index' Wikipedia page for the latest component list of the index, their FTSE industry classification benchmark sector and respective tickers. See Appendix 3 for the complete population. Dataset 2, the historical daily data in every node, is obtained from Yahoo Finance by indexing Dataset 1 tickers. The API requires an exact match with the ticker being looked-up; by adding a suffix of '.L' we solve a distinction with listings on the London Stock Exchange. We drop all variables except for Close from where we derive our features of interest declared earlier. We normalize and log-transform our close variable. Further, we annualize the log-transform (ALR) on a rolling window basis of a 1 week (ALR1W), 2 weeks (ALR2W), 1 month (ALR1M), and 2 months (ALR2M). Or 5, 10, 21 and 42 actual trading days. Finally, we derive the RSI and MACD.
The resulting 8 features vector attached to each node would be:

\[
H_{v}^{(0, t)}  = \left[ \mathrm{RSI}, \mathrm{MACD}, \mathrm{ALR1W}, \mathrm{ALR2W}, \mathrm{ALR1M}, \mathrm{ALR2M}, \mathrm{NormR}, \mathrm{LogR} \right]
\]

Dataset 3 is composed by a random selection of 28 fundamental ratios from the directory of 138 available in Yahoo Finance. In this study: currentRatio, quickRatio, profitMargins, grossMargins, operatingMargins, returnOnAssets, returnOnEquity, priceToBook, trailingPE, forwardPE, enterpriseToRevenue, enterpriseToEbitda, debtToEquity, revenueGrowth, earningsGrowth, revenuePerShare, payoutRatio, trailingAnnualDividendYield, fiveYearAvgDividendYield, 52WeekChange, beta, totalRevenue, totalDebt, totalCash, sharesOutstanding,
bookValue, trailingEps and forwardEps. We normalize and store for later use in the graph composition.

The selection process is outlined in Table 1. With an initial number of 475,700 observations across 100 companies listed in the London Stock Exchange for the period of 2007-2024. When we try to assemble our 3D features matrix in the shape of (stocks, features, timestamps) python finds that our array dimensions are inconsistent. A manual inspection of the data starting and ending periods highlights that some stocks have been publicly listed more recently than the rest, as an example, the firm Haleon plc with ticker HLN.L is a recent de-merger from the GSK Group; hence, the ticker lacked the past daily data necessary for the training of the model. We exclude all 17 companies without complete trading days for our chosen study period (80869 observations). Our initial Pearson correlation based on returns do not require further changes to the population. Nevertheless, on our proposed fundamental ratios graph composition; although not immediately apparent on our spearman correlations, an inspection on Yahoo Finance yielded that some companies had no fundamental ratios available; a further script to detect 0.00000 or missing data found 4 companies, which were removed (19.028 observations). Our array dimensions were still inconsistent, a small script detected 6 trading days missing in one or more of our remaining 79 companies, being removed too from the population (474 observations). Our final dataset for the main analysis comprises 79 companies, with 8 features across 375,329 observations. \\
\begin{table}[h]
\centering
\caption{Data processing}
\begin{tabular}{lrr}
\hline
\textbf{} & \textbf{Companies} & \textbf{Observations} \\
\hline
Initial population of FTSE100 companies from 2006–2024 & 100 & 475,700 \\
Less: companies without complete data for the period & (17) & (80,869) \\
Less: companies missing ratios from the database & (4) & (19,028) \\
Less: dates where at least 1 company had a missing day & — & (474) \\
\textbf{Number of observations in the final population} & \textbf{79} & \textbf{375,329} \\
\hline
\end{tabular}
\end{table}
\\
Appendix 4 shows that our data is comprised of companies from 45 different sectors based on the ICB classification system of the FTSE100. Some industries hold a bigger weight of observations, with Support services, Banks and Household goods at 6, 5 and 5 companies respectively. Appendix 5 graphically shows that our data is closely distributed with small trading days differences possibly due to calendar holiday days changing on a year-to-year basis.

\subsection{Pre-processing}
It is important to note that while the features of our nodes’ are values that evolve daily in each time step, our graph composition is declared at $t_0$; based on the chosen shared characteristics. This graph structure remains fixed for all remaining periods, or in simpler terms its composition and construction are set at the beginning of the study.\par
The first component of our neural network graph is based on the sector classification adjacency matrix from dataset
1. The adjacency matrix represents an existing connection of nodes (edges) by Boolean True if present; otherwise False; we use astype(int) to translate into 1/0s. We iterate over each pair of stocks to see if the sector is exactly identical to recognize the connection. Using NetworkX based on our NumPy array, we can visualize it (Appendix 6). Our second component follows the Pearson correlation of returns proposed by Yin et al., (2021) and Sun et al., (2020). However, we diverge by also testing our proposed fundamental ratios spearman correlation. Both are executed using a python script, making sure to remove self-correlation, as it could potentially create a path for noise in the model. See the graph in Appendix 7. We combine both graphs to leverage on two consequential characteristics that relate our nodes:
\par

\begin{figure}[htbp]
    \centering
    \begin{minipage}{0.41\linewidth}
        \centering
        \includegraphics[width=\linewidth]{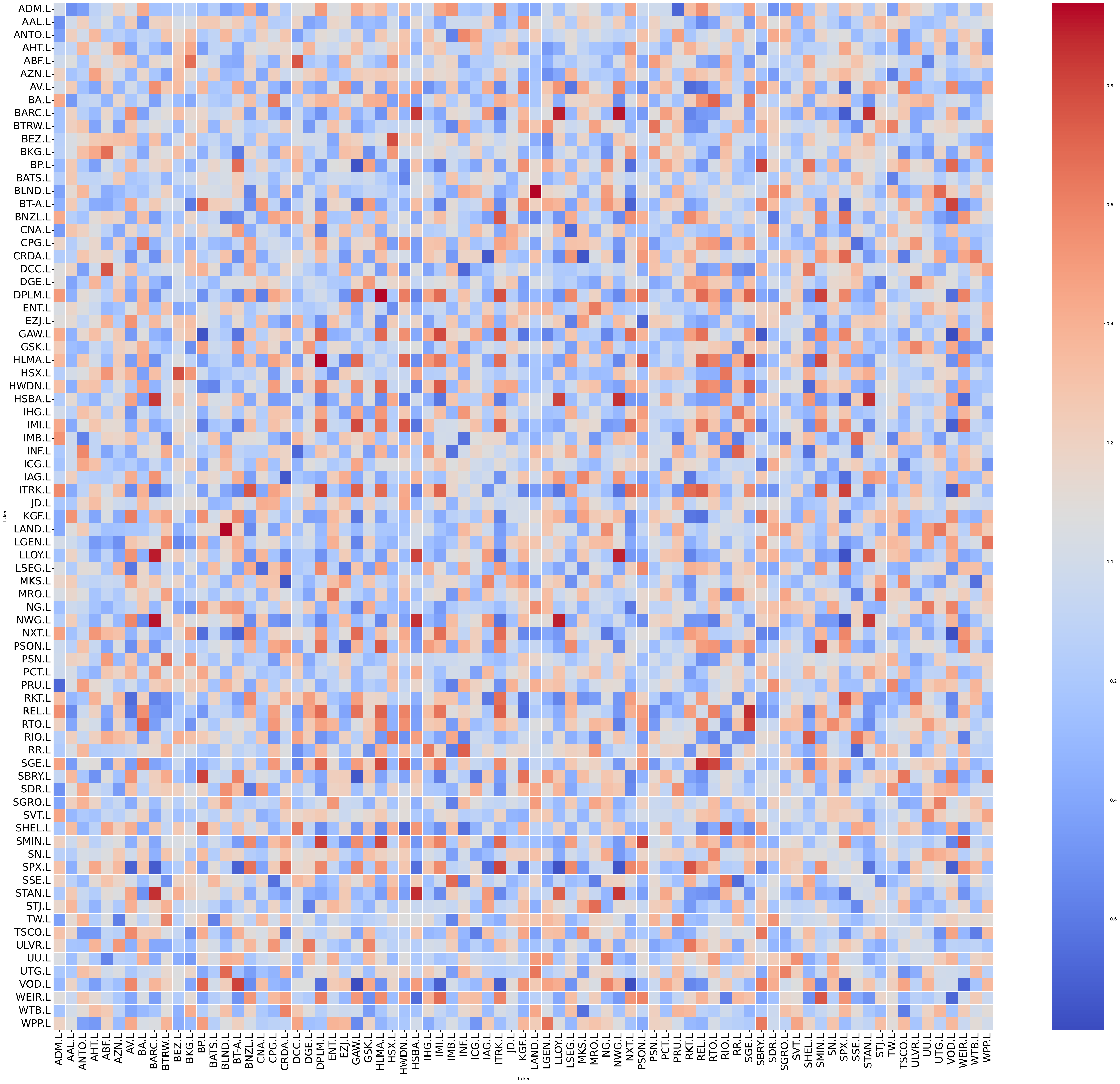}
    \end{minipage}
    \hspace{0.001\linewidth} 
    \begin{minipage}{0.48\linewidth}
        \centering
        \includegraphics[width=\linewidth]{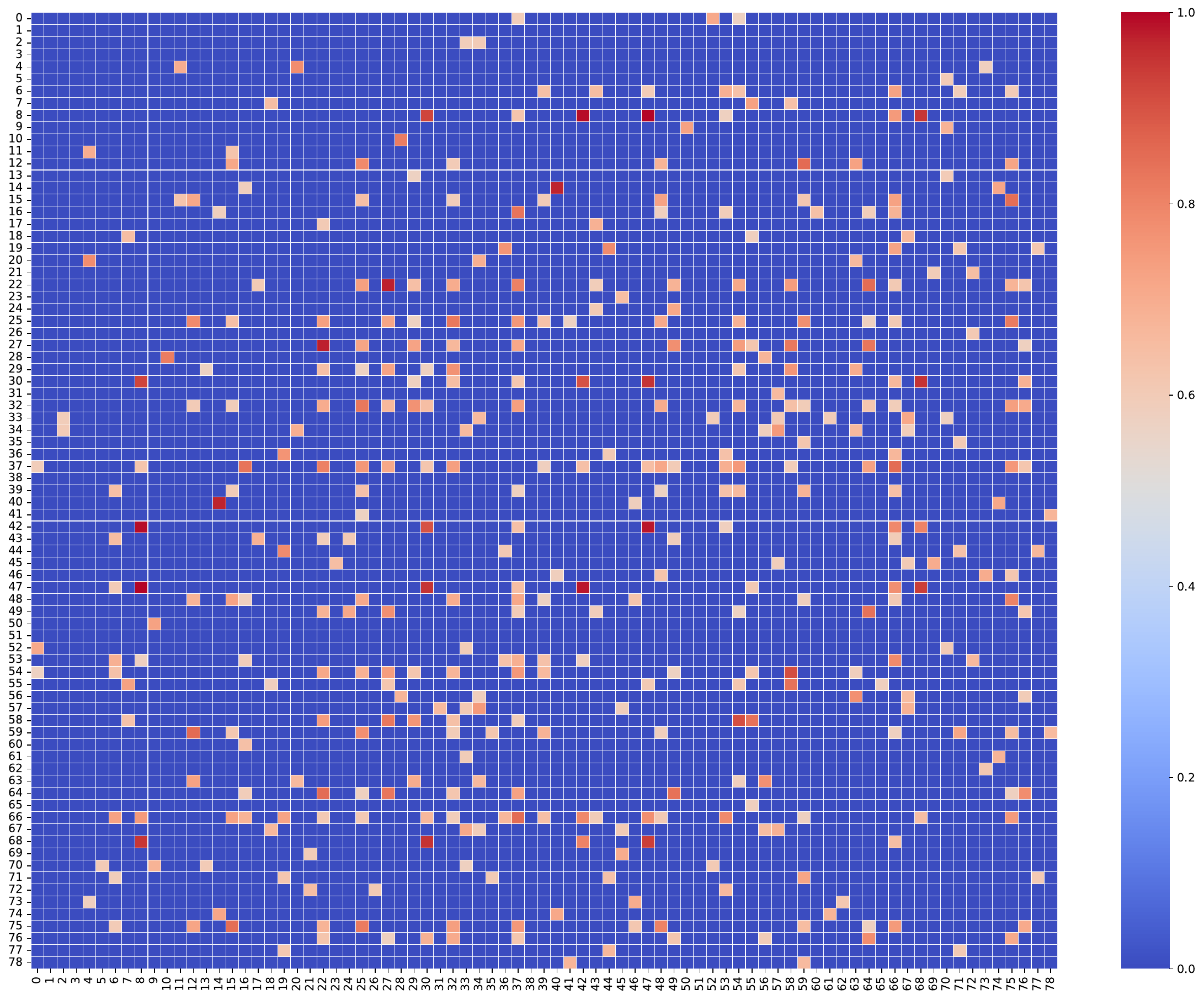}
    \end{minipage}
    
    \vspace{0.5em} 
    
    \begin{minipage}{0.41\linewidth}
        \centering
        \includegraphics[width=\linewidth]{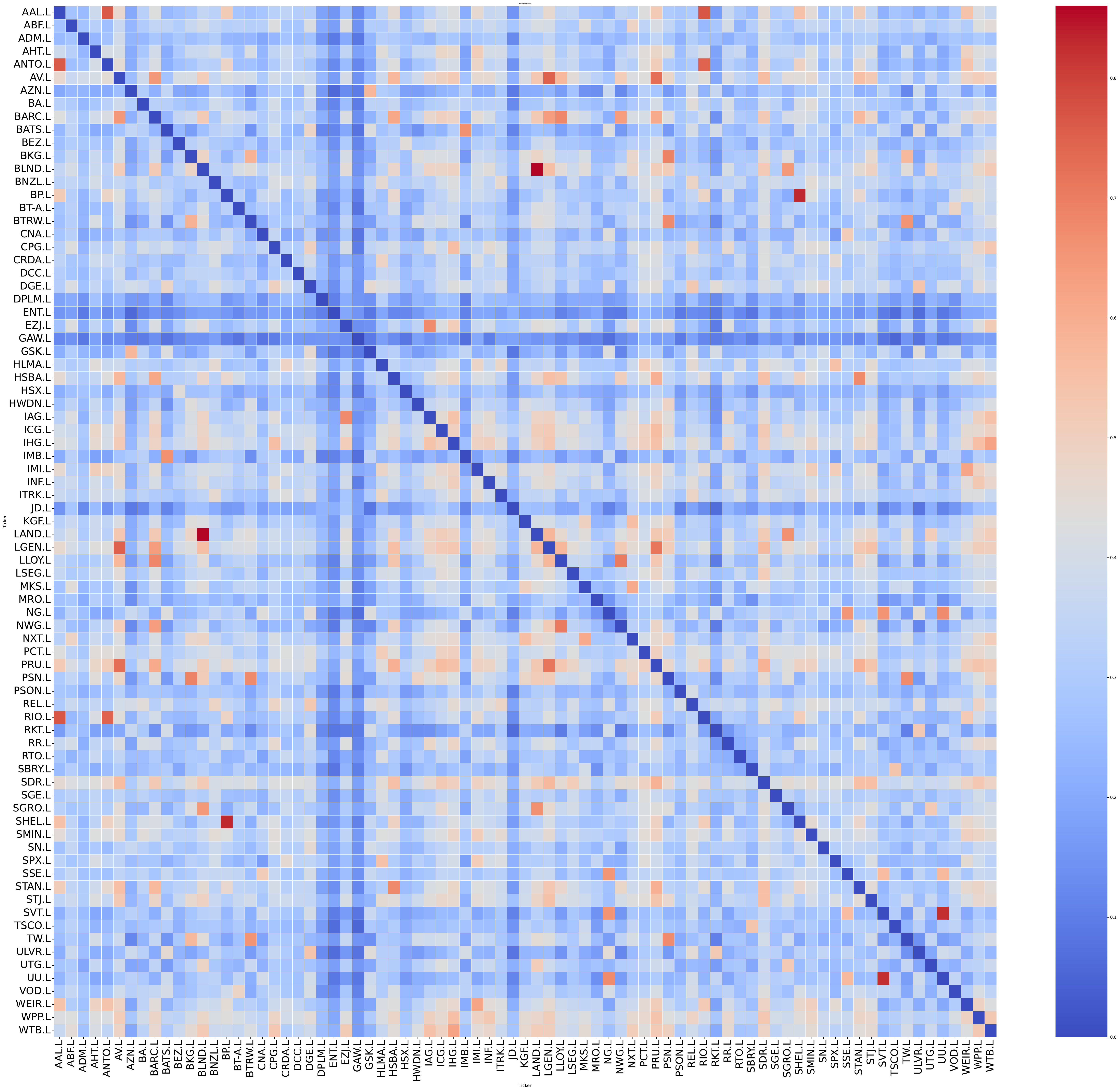}
    \end{minipage}
    \hspace{0.001\linewidth} 
    \begin{minipage}{0.48\linewidth}
        \centering
        \includegraphics[width=\linewidth]{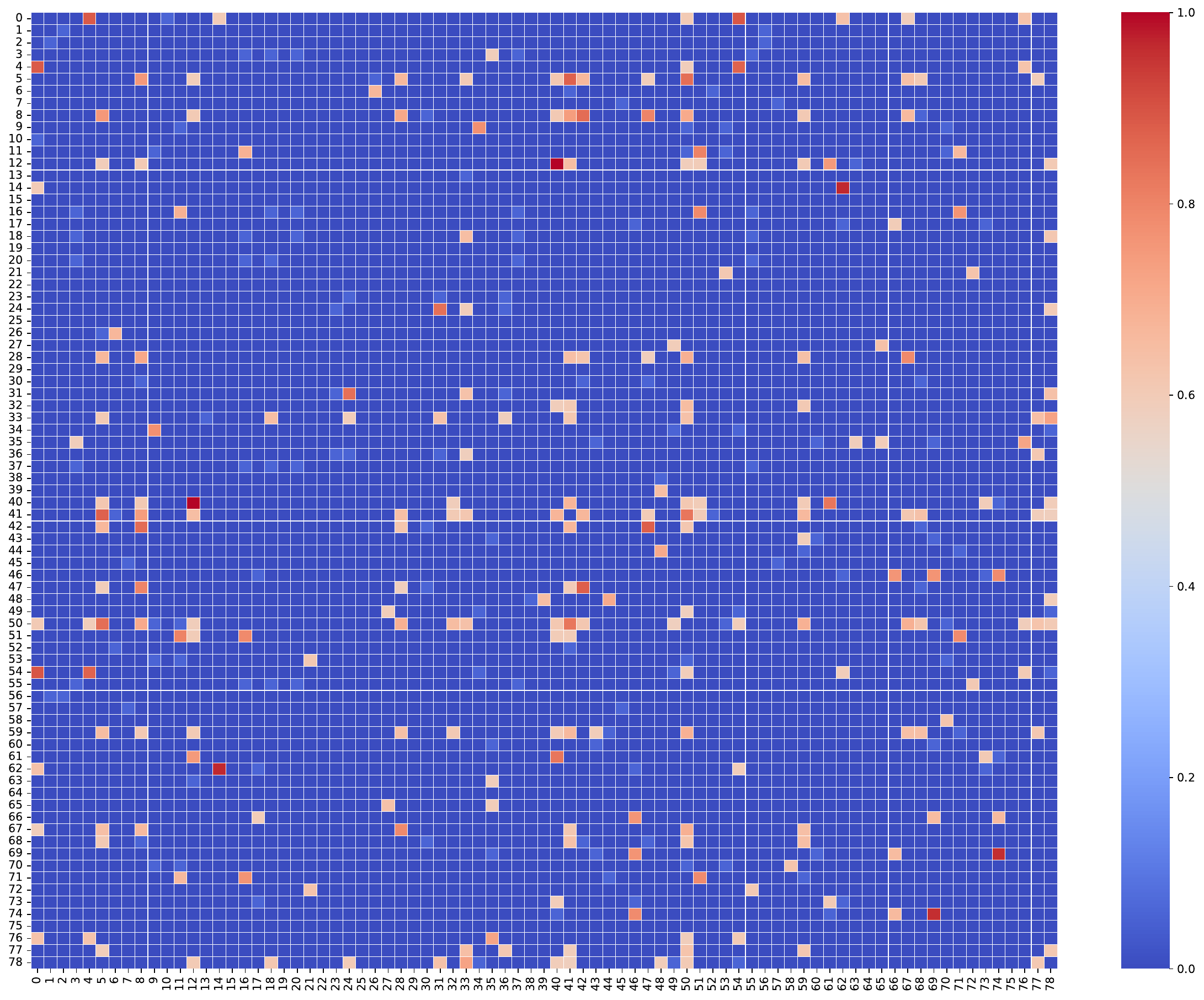}
    \end{minipage}
    
    \caption{Top, fundamental correlation based; bottom, returns correlation based}
    \label{fig:heatmap_grid}
\end{figure}

We achieve our final adjacency matrix by thresholding the correlations to the highest strength which minimizes too the number of unconnected nodes. 0.55 for the fundamental’s combination and 0.50 for the return’s correlation combination. See Appendix 8-9. We combine our features at each timestep with their respective node to create the data that will feed the training of our model. Our final processed data for our period of study has a size of 54.5GB.
\\

\subsection{Training}
Fundamentally our architecture processes time-varying graph data in two coupled phases: spatial feature extraction and temporal dynamics learning. This is achieved via two identical stacked TGCN cells which are individually composed of a GRU structure processing the output of a 2-layer GCNConv (See figure 2) in each instance. The first GCN layer aggregates the input graph (with our initial 8 features per node) in a 1-hop neighbor basis, learning local patterns of directly connected nodes with an output of 16 dimensions. The second GCN layer captures the 2-hop neighbor features, the global-level patterns, keeping the same dimensionality as the previous output (16, 16). The GRU components captures the first short-term history from the hidden states in its Update gate (line\_u), Reset gate (lin\_r) and Candidate state (lin\_c). The second TGCN cell further refines the 16-dimensional node embeddings.
However, the GRU gate now takes 48 inputs, suggesting an additional use of context. Finally, an attention mechanism (based on fixed neighbor weights) assigns importance weights to nodes or timesteps, enhancing focus on critical elements for the learning of the model. The architecture concludes with a linear output layer, reducing the 16 features to the required number of predictions.
\\
\setlength{\fboxsep}{0.1pt} 
\begin{figure}[ht]
    \centering
    \colorbox{black}{%
        \fbox{%
            \includegraphics[width=0.6\linewidth]{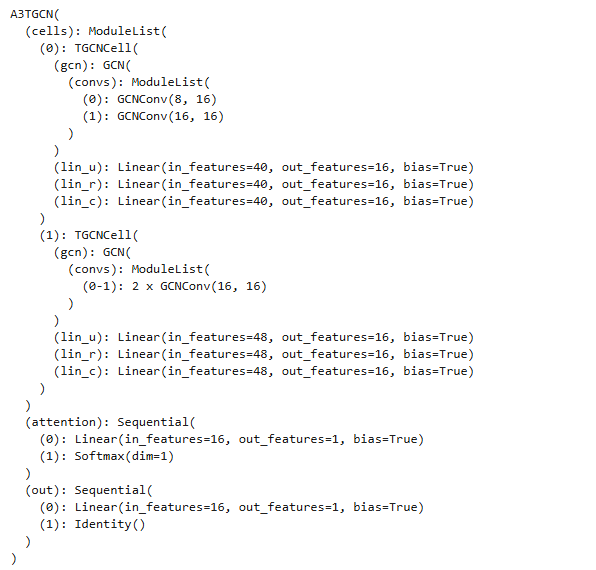}%
        }
    }
    \caption{The string representation of the A3TGCN model in PyTorch}
    \label{fig:placeholder}
\end{figure}
\\

\subsection{Hyperparameters tested}
Bang and Ryu (2023) found that longer predicting windows have lower accuracy but more meaningful learning, with the inverse trend as the window shortened. We test their conclusions by modifying the hyperparameters to create different versions of our model. Specifically, we increase the input length available for the next prediction from 5 to
30. We then further differentiate the versions by testing the robustness of the model; as we depart further from the immediacy of the sequence length. We increase the future window prediction to double, triple and eight times further in the future. We use a standard configuration of 0.005 learning rate, 0.00001 weight decay, 32 batch size, and 10 epochs due to computational constraints.

\begin{table}[H]
\centering
\caption{Parameters configurations to be tested}
\begin{tabular}{c c c}
\hline 
Past window & Future window & ID \\
\hline
\multirow{4}{*}{5 sequence length} & 1d & Version 1 (5SL1D) \\
 & 2d & Version 2 (5SL2D) \\
 & 3d & Version 3 (5SL3D) \\
 & 8d & Version 4 (5SL8D) \\
\hline
\multirow{4}{*}{30 sequence length} & 1d & Version 5 (30SL1D) \\
 & 2d & Version 6 (30SL2D) \\
 & 3d & Version 7 (30SL3D) \\
 & 8d & Version 8 (30SL8D) \\
\hline
\end{tabular}
\label{tab:parameters}
\end{table}

\subsection{Evaluation criteria}
Evaluation metrics are used to measure the accuracy and effectiveness of the model in predicting the actual labels for the period. Per existing literature, Patel et al (2024) observes MAE and RMSE as the most commonly used metrics to evaluate the regression task of model; i.e. predicting continues values such as future closing prices.
Nevertheless, work by Shah et al. (2022) report many other methods in use with no clear consensus in the field of study. Under these circumstances, in a bid to enhance future comparability; MSE and MRE are included as well. See Appendix 11 for the equations.

\section{Results}
We first run a one-sample t-test to choose the best graph composition. The results in Table 3 show that the Pearson correlation on returns outperforms our fundamental ratios Spearman correlation approach, and thus we follow that approach for our results. We reject our null hypothesis \( H_0: \mu_d = 0 \) as the highly significant (\( p = 1.110\times10^{-76} \)) negative t-statistic (\( t = -18.5332 \)) provides evidence to reject it in favor of the left-tailed alternative \( H_A: \mu_d < 0 \); i.e., the squared error for Pearson-based predictions is on average lower than for Spearman-based ones.
\\
\begin{table}[H]
\centering
\caption{One-sample t-test results}
\begin{tabular}{c c c c c}
\hline
Mean & T-test & P-value & \multicolumn{2}{c}{95\% Confidence Interval} \\
\hline
-0.0008468 & -18.5332 & $<0.00001$ & -0.0009364 & -0.0007573 \\
\hline
\multicolumn{5}{l}{$H_0$: mean = 0} \\
\multicolumn{5}{l}{$H_A$: mean $<$ 0} \\
\hline
\end{tabular}
\label{tab:table3}
\end{table}

As previously noted, GNNs are affected under the black box property. Even though we are unable to locate the importance of a feature in the general learning of our model, our testing draws some interesting findings. To begin with, the Primary case performs best across all metrics. Its low MSE value (0.0044) indicates large errors are infrequent in the model; since these are given more weight in that metric calculation. However, the RMSE (0.0660) being higher than MAE (0.0441) signals they are still present. On average the MRE metrics highlights that the model is off by 4.17\% of the actual stock price; i.e. allowing for price scale differences to be disregarded. A trend is found between future window size and the different error metrics. As the model is forced to predict further in the future, its accuracy diminishes by almost triple when comparing the 5SL1D to 5SL8D under MAE and RMSE, double in MRE and decreases significantly nine times for MSE. Yet, the increase in sequence length (from 5 to 30) mitigates the rate at which errors grow as the prediction horizon extends. On an immediate next day prediction, the model is worse when given more input (MAE 0.0441/0.0512); nevertheless, as we drive the model towards tougher further in the future predictions more extensive historical context provides a smaller marginal incremental error; with the 30SL variation offering lower errors across all metrics starting from 3 days future window.

\begin{table}[H]
\centering
\caption{Error metrics of the tested versions}
\begin{tabular}{c c c c c}
\hline
Version & MAE & MSE & RMSE & MRE \\
\hline
Primary & 0.0441 & 0.0044 & 0.0660 & 3.46\% \\
Version 2 (5SL2D) & 0.0551 & 0.0057 & 0.0753 & 5.23\% \\
Version 3 (5SL3D) & 0.0891 & 0.0141 & 0.1186 & 8.47\% \\
Version 4 (5SL8D) & 0.1227 & 0.0288 & 0.1698 & 11.67\% \\
Version 5 (30SL1D) & 0.0512 & 0.0051 & 0.0712 & 4.86\% \\
Version 6 (30SL2D) & 0.0560 & 0.0061 & 0.0783 & 5.32\% \\
Version 7 (30SL3D) & 0.0574 & 0.0065 & 0.0808 & 5.46\% \\
Version 8 (30SL8D) & 0.0850 & 0.0139 & 0.1179 & 8.09\% \\
\hline
\end{tabular}
\label{tab:table4}
\end{table}

\par
We poise that our ALR features provide significant context information even in short sequence lengths. Their nature allows them to contain long term information (5-42 days) in a compact data point. By removing these context features we find that a shorter sequence (5 previous days) no longer retains the smaller error advantage over a longer sequence (30), see Table 5. We establish that our ALR features are more computing efficient, by preventing the need to increase the sequence length. It should be noted that their use increases on average 3:03 minutes of computing time as the model need to parse through the new data, yet this is a lesser amount than the increase of 6:48 minutes when using a 30-sequence length with no ALR features. Allowing for better performance under a constraint of computing power. Moreover, our initial findings about the robustness of errors when increasing the future window through a longer sequence length still hold; however, we contend that the stochastic nature of our model introduces volatility in our results requiring re-testing to ensure the small margin of difference is not invalidated in the possible confidence interval of our distribution errors.\\

\begin{figure}[H]
    \centering
    \includegraphics[width=1\linewidth]{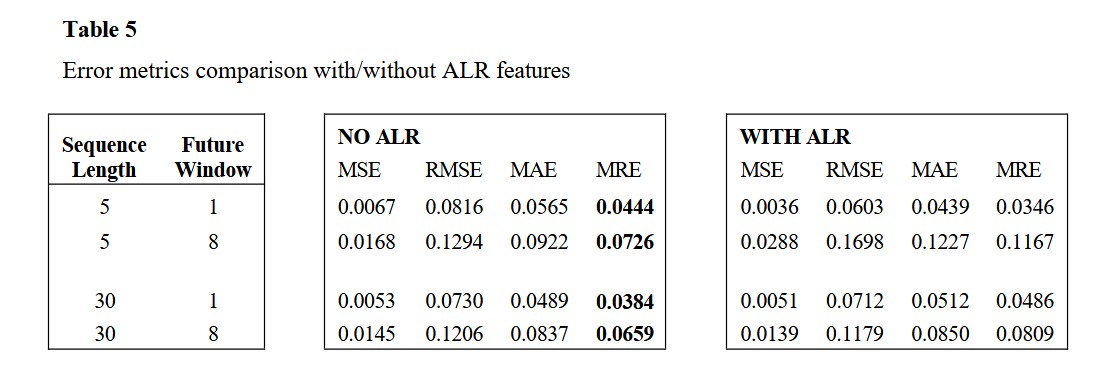}
  
\end{figure}

The specific categorization of 25 industries in the LSE combined with a medium strength correlation threshold result in some of our nodes remaining unconnected. Contrarian to the findings of Zanardi and Serrano (2024), who on individual GCN and GAT architectures found isolated nodes unable to learn effectively, we do not observe a noticeable degradation in performance attributable to unconnected nodes; further the low MSE in all versions points towards the absence of unpredictable nodes, our plots support the idea that the model presents no apparent issue for unconnected nodes, see Appendix 12. We posit that the vast amount of data offered (30008 pieces of data = 4751 daily observations * 8 features) to this specific architecture is enough to generalize learned parameters from the connected nodes towards an unconnected.\par
Further evaluation of individual industries is necessary. The error metrics presented are related to the 79 predicted companies at the same time. A visual inspection of the predicted outputs in our plots (Appendix 12) shows there are stocks where the model performs relatively worse; a case-by-case basis evaluation could introduce improved results if the model is isolated in use only for specific industries. It should be noted that non-performing nodes should not be removed if proceeded as recommended, as they still provide context for the learning of the model used in its more accurate nodes.\par
Due to computational constraints, we instead estimated a learning curve on a smaller dataset (5 years), see Figure 4. Nakkiran et al. (2019) showed the double descent phenomenon implies that analysing model behaviour with smaller datasets can still yield valuable insights. We retest our Primary as it was the case producing the lowest error under 56 epochs (minimum validation error point) obtaining MSE: 0.0052, RMSE: 0.0722, MAE: 0.0466, MRE: 0.0366. Compared to our previous results in Table 4, a reduction of -0.0023 MSE, -0.0203 RMSE, -0.0140 MAE and an increase of 0.0038 MRE. The decrease in absolute error metrics (MSE, RMSE, MAE) and the increase in MRE suggest the model accuracy improvement under higher training iterations at the cost of large relative errors as the MRE is more sensitive to small absolute errors. Therefore, we conclude that the model is likely predicting higher- scale stocks better.
\begin{figure}[H]
    \centering
    \includegraphics[width=0.75\linewidth]{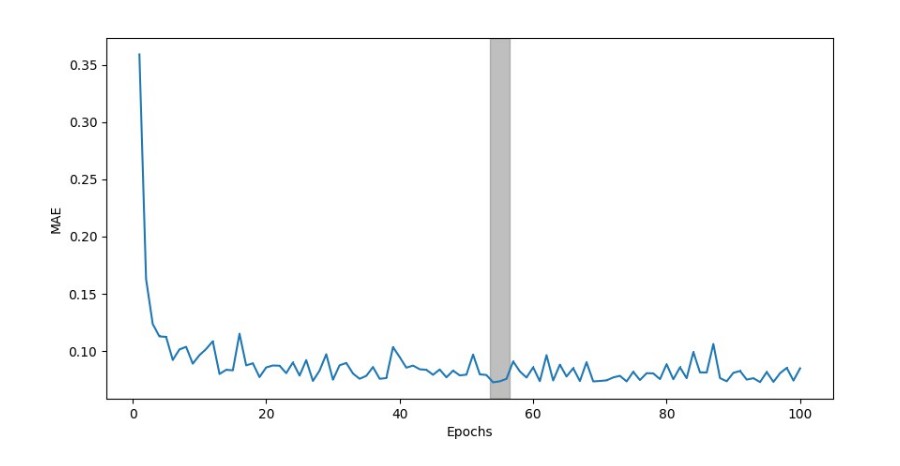}
    \caption{Training loss curve, MAE vs Epochs}
    \label{fig:placeholder}
\end{figure}
\section{Conclusion}
This study purpose was to test predicting stock prices in a previously proposed neural network for a different task. Where our results indicate that the model is successfully able to track the pattern of change of stocks (R2 = 0.9936105) with small deviation in actual values or low absolute errors via different metrics (MSE: 0.0036, RMSE: 0.0603, MAE: 0.0439) in our Primary case. We further explored different changes in configurations to better understand and maximize the potential of the architecture.\par
Our results were both contradictory and consistent with our expectations. We anticipated our Spearman correlation on fundamental ratios to play a deeper role in improving predictive power; however, contradictorily, our tests and results indicated that there was not a noticeable improvement but rather an underperformance when compared to the previous literature which supported the use of Pearson correlation based on historical returns. This highlights how little of a role graph composition played in our setup, signifying that efforts should be aligned toward the features selection, their pre-processing and the configuration of hyper-parameters; as long as most nodes appear connected. A possible explanation could be related to the fact that a momentum-based model propagating return-related features would mostly benefit from nodes connections based on returns correlations too. We hoped the model would gain from the nuances of fundamental ratios usually used by investors to understand the current financial-operational position of a company against its historical performance and compared to industry peers. However, it is plausible to believe this strategy required more structural factors, with lower frequency of rebalancing, where some fundamental ratios (especially if chosen randomly) do not particularly represent trends that manifest in the long run.\par
Nevertheless, our annualized log returns features (ALR) were consistent with our expectations in reducing the computing time required. We predicted that a feature containing long term information would eliminate the need to stack multiple trading day’s pre-processed data-graphs; thus, reducing the computational burden and our results supported the notion by achieving closely similar error metrics under less computing time by using ALR with a short sequence length. However, we did not expect that the increase in the number of features itself would dramatically increase the time too by 50\% or 3 minutes on average per training/testing cycle. We initially noticed our processed complete data weighted 30.1GB during early trials, with a single data-graph weighting 7017KB, we estimated that an additional excel cell (an extra feature for 1 trading day in pre-processed data) would not be equally close in size, which lead us to add 4 more ALR features. This was not an equitable comparison as when processed, every feature added 1401KB to every compatible data-graph increasing the total size to 54.5GB or 12622KB individually. Still 30 stacked old data-graph still would weight more than 5 of our new heavier data-graph which explains why our computing time was faster. The key conclusion remains that the use of ALR features effectively streamlined the training process without negatively impacting model performance, making the increase in processing time a secondary concern relative to the broader goal.\par
On related research, part of our methodology was supported by the applications of Sun et al. (2020). We achieve improved results in 0.01308 less RMSE and 0.00296 less MAE against their proposed MCT-GCN architecture. However, since their results follow a different dataset, sample period and model configuration a direct comparison making the contrast ofnot straightforward. However, it remains one of the only comparable works in the area and can still serve as a benchmark that our architecture offers better predictive power.\par
The primary limitation of this study is the model's extensive capacity. Although not immediately apparent, the model possesses the potential to incorporate a wide range of variations and extensions. For example, future work could explore alternative graph compositions, such as a value chain graph that connects companies through client/customer relationships, similar to the SPLC Bloomberg function. Additionally, the model can incorporate endless different features. Moreover, the necessity for hyperparameter fine-tuning in each case remains a significant challenge as conceptualizing and computing all these scenarios represents a herculean task. As a result, despite our best efforts, the true optimal scenario for stock prediction remains uncertain. Future research, on a more immediate relation to our methodology could test longer ALR features and reduce their number to only two. Testing could be done on individually selecting fundamental ratios for the graph composition that best represent longer term relationships.\par
On several other limitations: On our chosen population, the FTSE 100 index updates its constituent members on a quarterly basis to reflect the top market capitalized businesses. Furthermore, some listed companies lack historical data extending beyond our cutoff date of 2007, particularly if their IPO date occurred more recently. Consequently, survivorship bias is present for this study. Businesses that failed are not part of the data being analyzed and we mostly deal with mature companies that have been listed for an extended period of time. On real-world applications, our best results of 3.46\% mean relative error suggests that this strategy might be less profitable for mature blue-chip stocks, where scale sensitivity makes even modest errors significant. A better interaction platform could be found in more volatile environments, such as small-caps, where the pertinent predicted information can still offer alpha after accounting for this inaccuracy expense. On specific model limitations, we point out that this model is inherently limited against event-driven price movements. Since it operates on a momentum basis from the learning of price trends, with no access to relevant exterior information, it is highly unlikely to predict an earnings report or the announcement of a M\&A deal; unless insider activity influences price discovery via bid-ask operations in the days leading to the announcement. On real-world relevance, Fischer and Krauss (2018) noted the LSTM strategy to have been arbitraged away after 2010; with profitability around zero after transactions costs. Market adaptation would ensure that as this strategy continues to be studied, other market participants could exploit it, reducing its effectiveness over time. Nevertheless, the continuing expanding family of GNN architectures, even if explored towards different uses, could allow for a longer resilience to arbitrage pressures due to the small nuances in-between architectures which would result in different buy/sell signals.\par 

\pagebreak 

\section {Appendix}
Appendix 1: Graph Neural Network classification tree, blue square denotates our model, adapted and edited from Zhou et al. (2020)

\begin{figure}[H]
    \centering
    \includegraphics[width=0.8\linewidth]{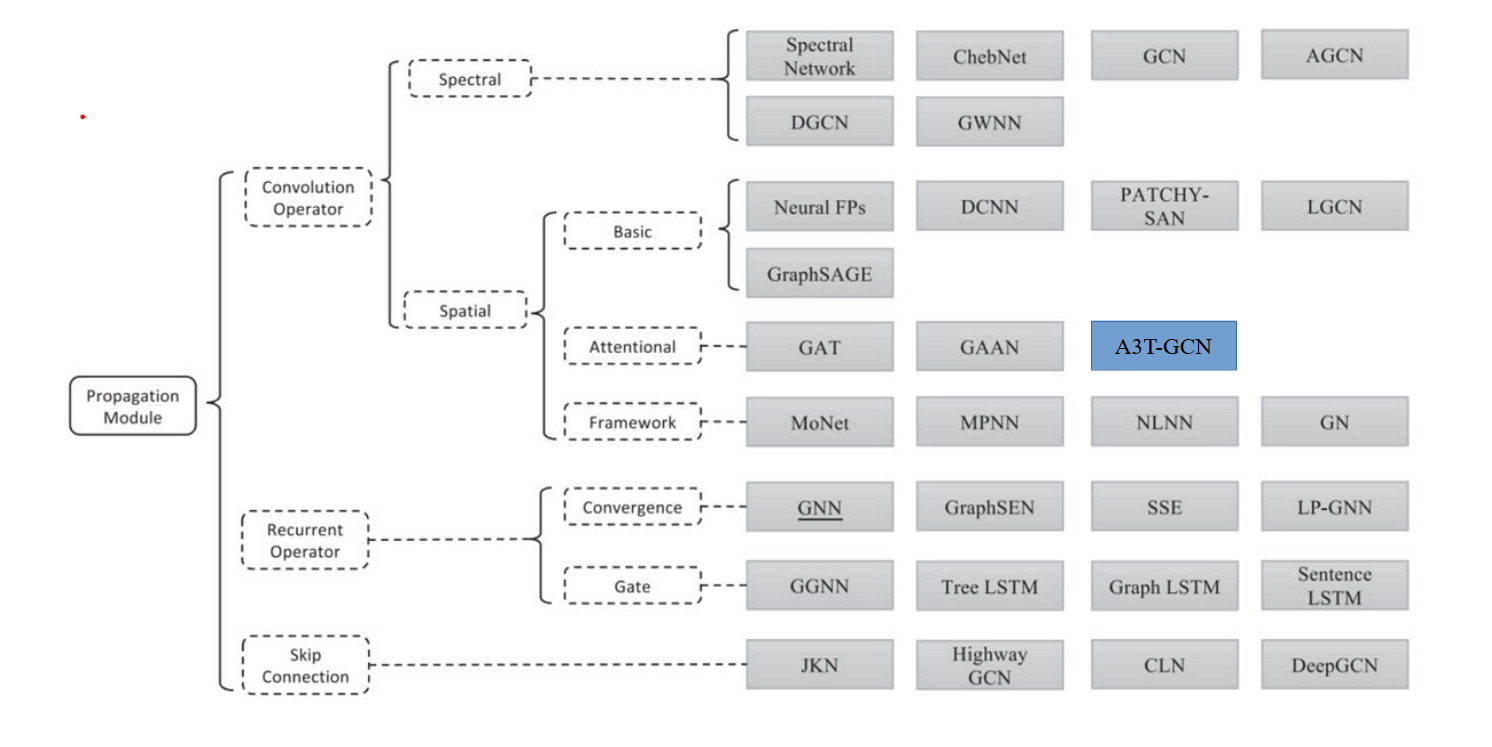}
    \label{fig:placeholder}
\end{figure}

\noindent Appendix 2: A3T-GCN cell overview, adapted from (Zhu et al., 2020)

\begin{figure}[H]
    \centering
    \includegraphics[width=0.8\linewidth]{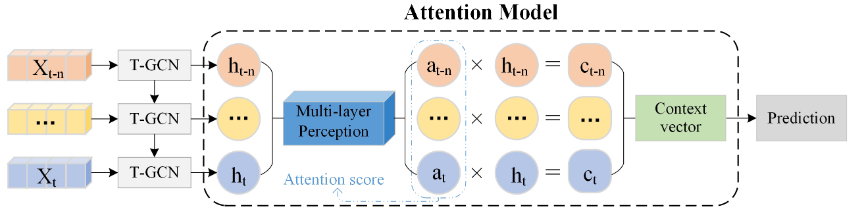}
    \label{fig:placeholder}
\end{figure}

\pagebreak
\noindent Appendix 3: Sample universe after selection process
\\
\begin{longtable}{llp{8cm}l}
\toprule
Ticker & Company Name & FTSE Industry Classification Benchmark Sector \\
\midrule
A3T-GCN & Admiral Group & Insurance \\
AAL.L & Anglo American plc & Mining \\
ANTO.L & Antofagasta plc & Mining \\
AHT.L & Ashtead Group & Support services \\
ABF.L & Associated British Foods & Food \& tobacco \\
AZN.L & AstraZeneca & Pharmaceuticals \& biotechnology \\
AV.L & Aviva & Life insurance \\
BA.L & BAE Systems & Aerospace \& defence \\
BARC.L & Barclays & Banks \\
BTRW.L & Barratt Redrow & Household goods \& home construction \\
BEZ.L & Beazley & Insurance \\
BKG.L & Berkeley Group Holdings & Household goods \& home construction \\
BP.L & BP & Oil \& gas producers \\
BATS.L & British American Tobacco & Tobacco \\
BLND.L & British Land & Real estate \\
BT-A.L & BT Group & Telecommunications services \\
BNZL.L & Bunzl & Support services \\
CNA.L & Centrica & Multiline utilities \\
CPG.L & Compass Group & Support services \\
CRDA.L & Croda International & Chemicals \\
DCC.L & DCC plc & Support services \\
DGE.L & Diageo & Beverages \\
DPLM.L & Diploma & Industrial Support services \\
ENT.L & Entain & Travel \& leisure \\
EZJ.L & EasyJet & Travel \& leisure \\
GAW.L & Games Workshop & Leisure Goods \\
GSK.L & GSK plc & Pharmaceuticals \& biotechnology \\
HLMA.L & Halma plc & Electronic equipment \& parts \\
HSX.L & Hiscox & Non-life Insurance \\
HWDN.L & Howdens Joinery & Homebuilding \& construction supplies \\
HSBA.L & HSBC & Banks \\
IHG.L & IHG Hotels \& Resorts & Travel \& leisure \\
IMI.L & IMI & Machinery, tools, heavy vehicles, trains \& ships \\
IMB.L & Imperial Brands & Tobacco \\
INF.L & Informa & Media \\
ICG.L & Intermediate Capital Group & Financial services \\
IAG.L & International Airlines Group & Travel \& leisure \\
ITRK.L & Intertek & Support services \\
JD.L & JD Sports & General retailers \\
KGF.L & Kingfisher plc & Retailers \\
LAND.L & Land Securities & Real estate investment trusts \\
LGEN.L & Legal \& General & Life insurance \\
LLOY.L & Lloyds Banking Group & Banks \\
LSEG.L & London Stock Exchange Group & Financial services \\
MKS.L & Marks \& Spencer & Food \& drug retailing \\
MRO.L & Melrose Industries & Aerospace \& defence \\
NG.L & National Grid plc & Multiline utilities \\
NWG.L & NatWest Group & Banks \\
NXT.L & Next plc & General retailers \\
PSON.L & Pearson plc & Media \\
PSN.L & Persimmon & Household goods \& home construction \\
PCT.L & Polar Capital Technology Trust & Investment trusts \\
PRU.L & Prudential plc & Life insurance \\
RKT.L & Reckitt & Household goods \& home construction \\
REL.L & RELX & Media \\
RTO.L & Rentokil Initial & Support services \\
RIO.L & Rio Tinto & Mining \\
RR.L & Rolls-Royce Holdings & Aerospace \& defence \\
SGE.L & Sage Group & Software \& computer services \\
SBRY.L & Sainsbury's & Food \& drug retailing \\
SDR.L & Schroders & Financial services \\
SGRO.L & Segro & Real estate investment trusts \\
SVT.L & Severn Trent & Multiline utilities \\
SHEL.L & Shell plc & Oil \& gas producers \\
SMIN.L & Smiths Group & General industrials \\
SN.L & Smith \& Nephew & Health care equipment \& supplies \\
SPX.L & Spirax Group & Industrial engineering \\
SSE.L & SSE plc & Electrical utilities \& independent power producers \\
STAN.L & Standard Chartered & Banks \\
STJ.L & St. James's Place & Financial services \\
TW.L & Taylor Wimpey & Household goods \& home construction \\
TSCO.L & Tesco & Food \& drug retailing \\
ULVR.L & Unilever & Personal goods \\
UU.L & United Utilities & Multiline utilities \\
UTG.L & Unite Group & Real estate investment trusts \\
VOD.L & Vodafone Group & Mobile telecommunications \\
WEIR.L & Weir Group & Industrial goods and services \\
WTB.L & Whitbread & Retail hospitality \\
WPP.L & WPP & Media \\
\end{longtable}
\pagebreak

\noindent Appendix 4: Distribution of observations by FTSE industry classification benchmark (ICB) \& by trading days in period (Yearly)

\begin{figure}[H]
    \centering
    \includegraphics[width=1\linewidth]{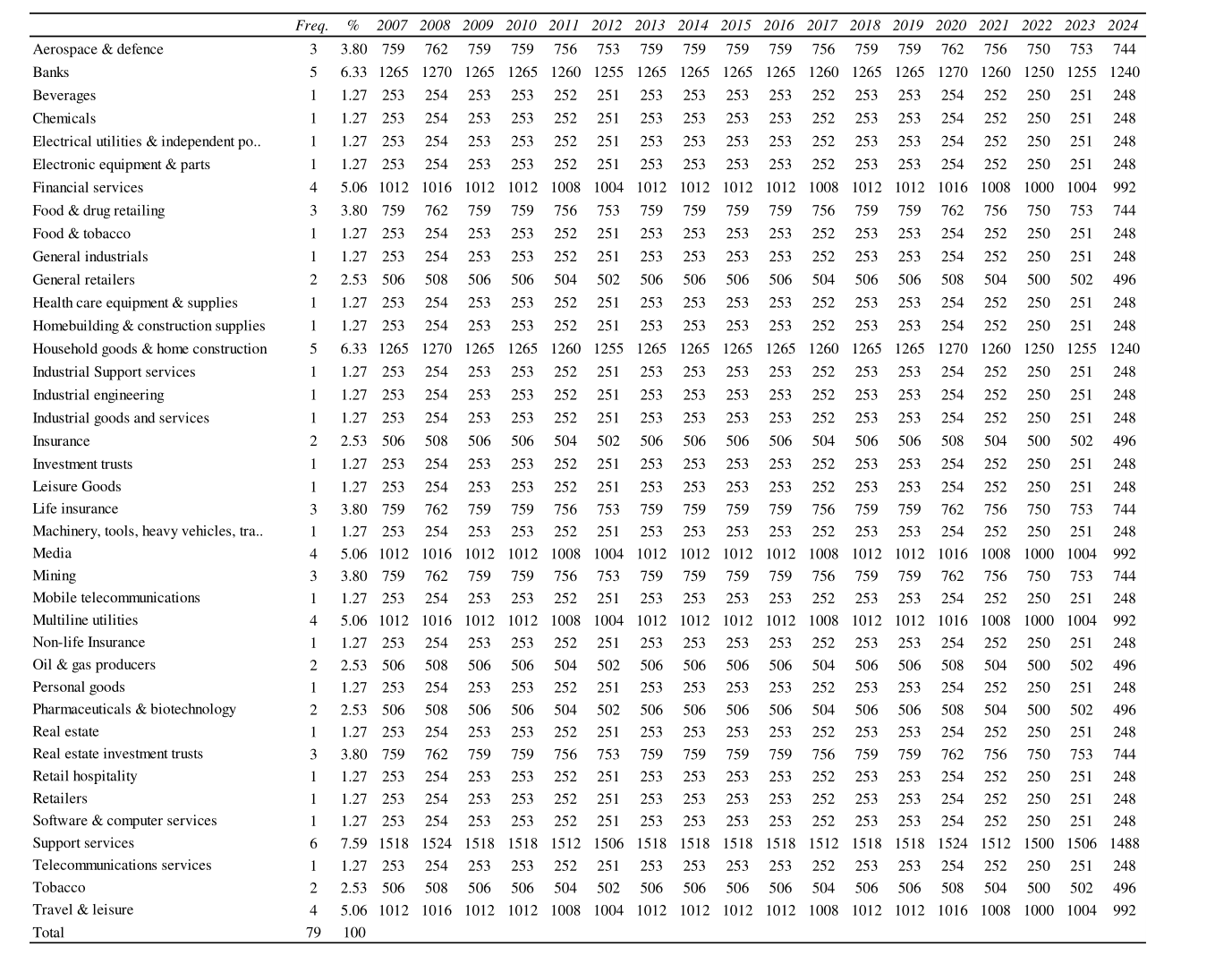}
    \label{fig:placeholder}
\end{figure}

\noindent Appendix 5: Total Sample of Trading Day Observations per Year

    \begin{figure}[H]
        \centering
        \includegraphics[width=0.8\linewidth]{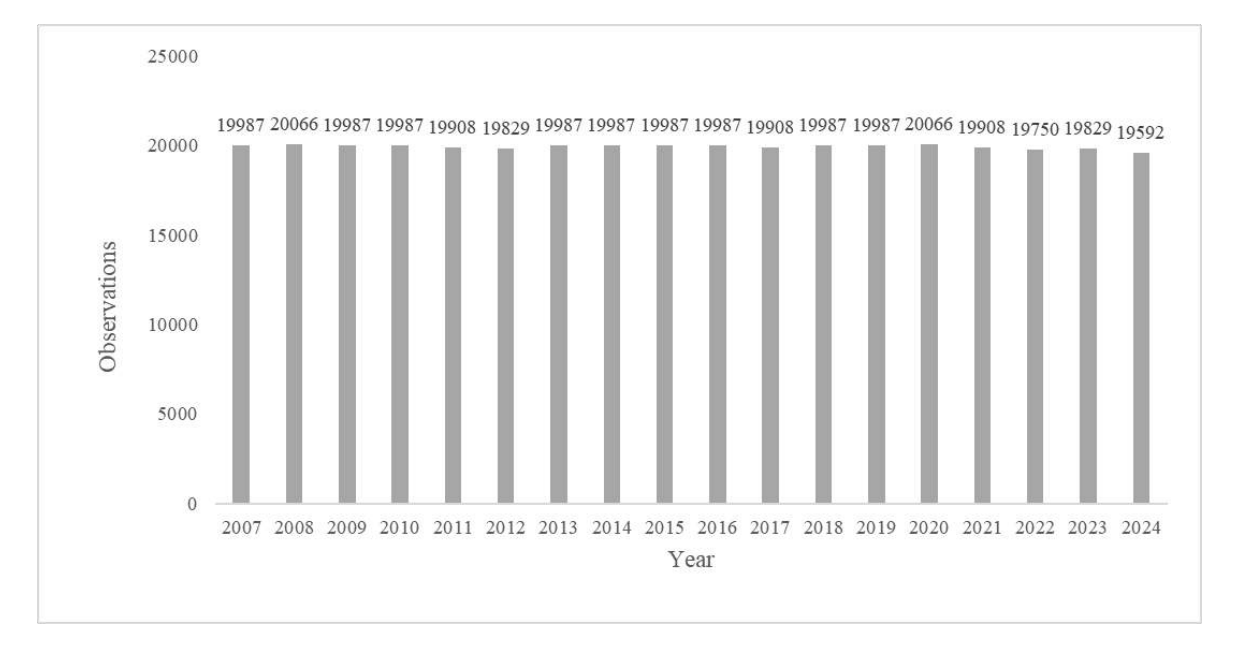}
        \label{fig:placeholder}
    \end{figure}
\pagebreak

\noindent Appendix 6: Nodes and edges graph by sector
\begin{figure}[H]
    \centering
    \includegraphics[width=0.75\linewidth]{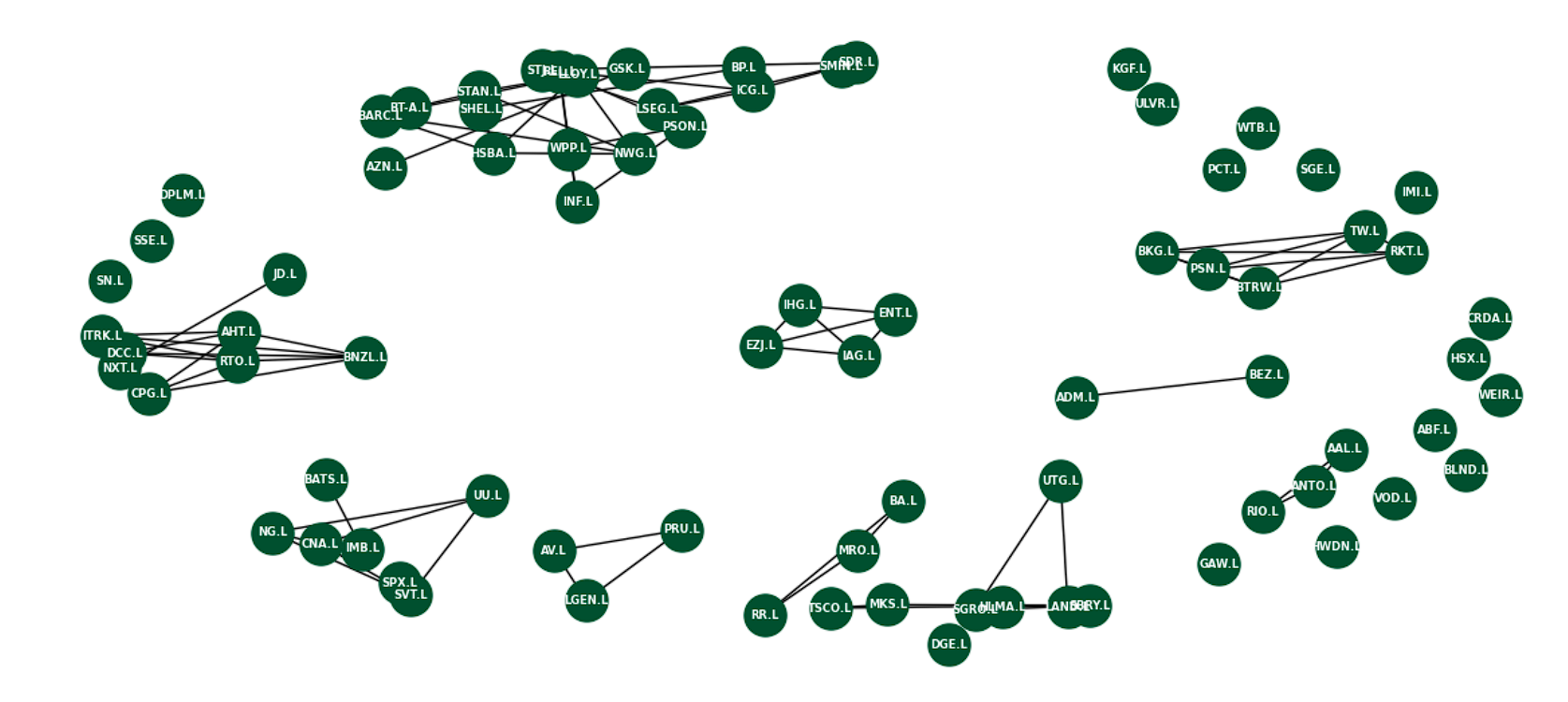}
    \label{fig:placeholder}
\end{figure}
\noindent Appendix 7: Nodes and edges graph by fundamental correlation
\begin{figure}[H]
    \centering
    \includegraphics[width=0.75\linewidth]{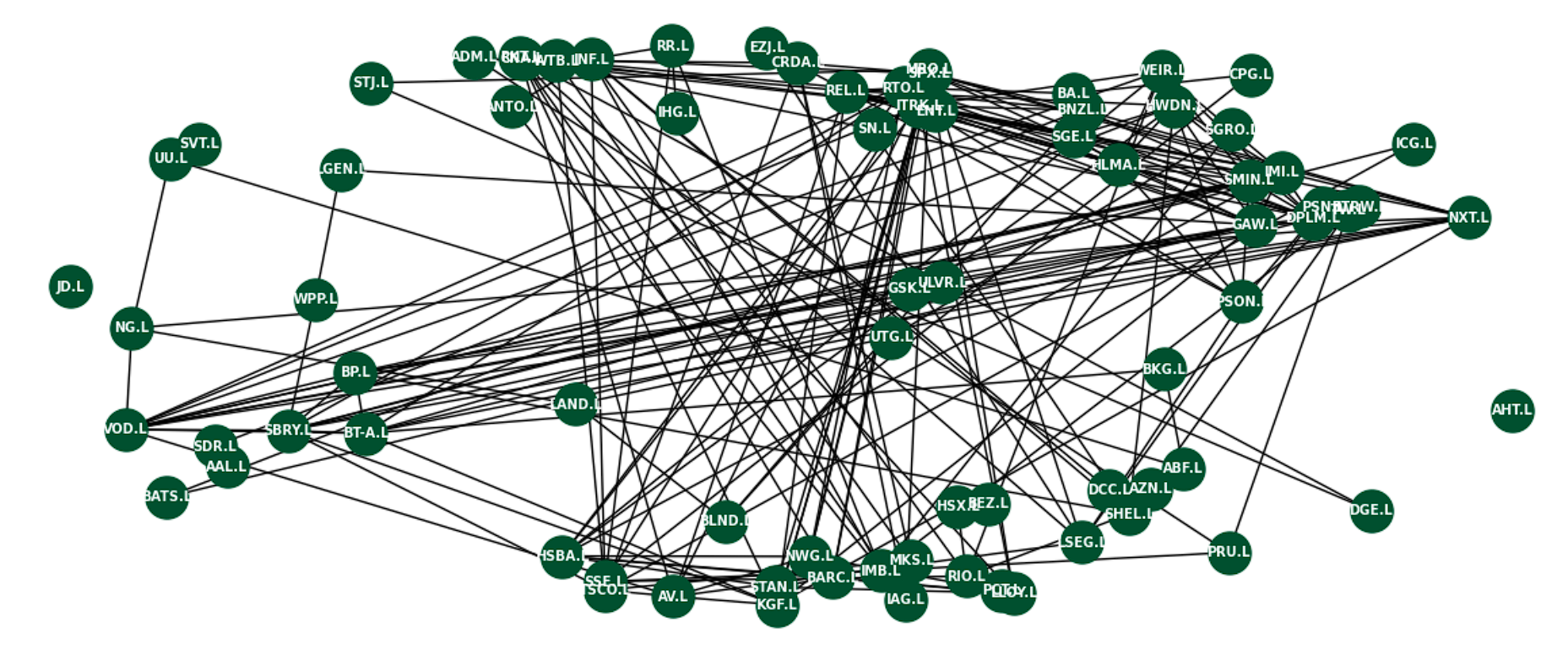}
    \label{fig:placeholder}
\end{figure}
\noindent Appendix 8: Nodes and edges graphs by sector and Pearson correlation of returns
\begin{figure}[H]
    \centering
    \includegraphics[width=0.75\linewidth]{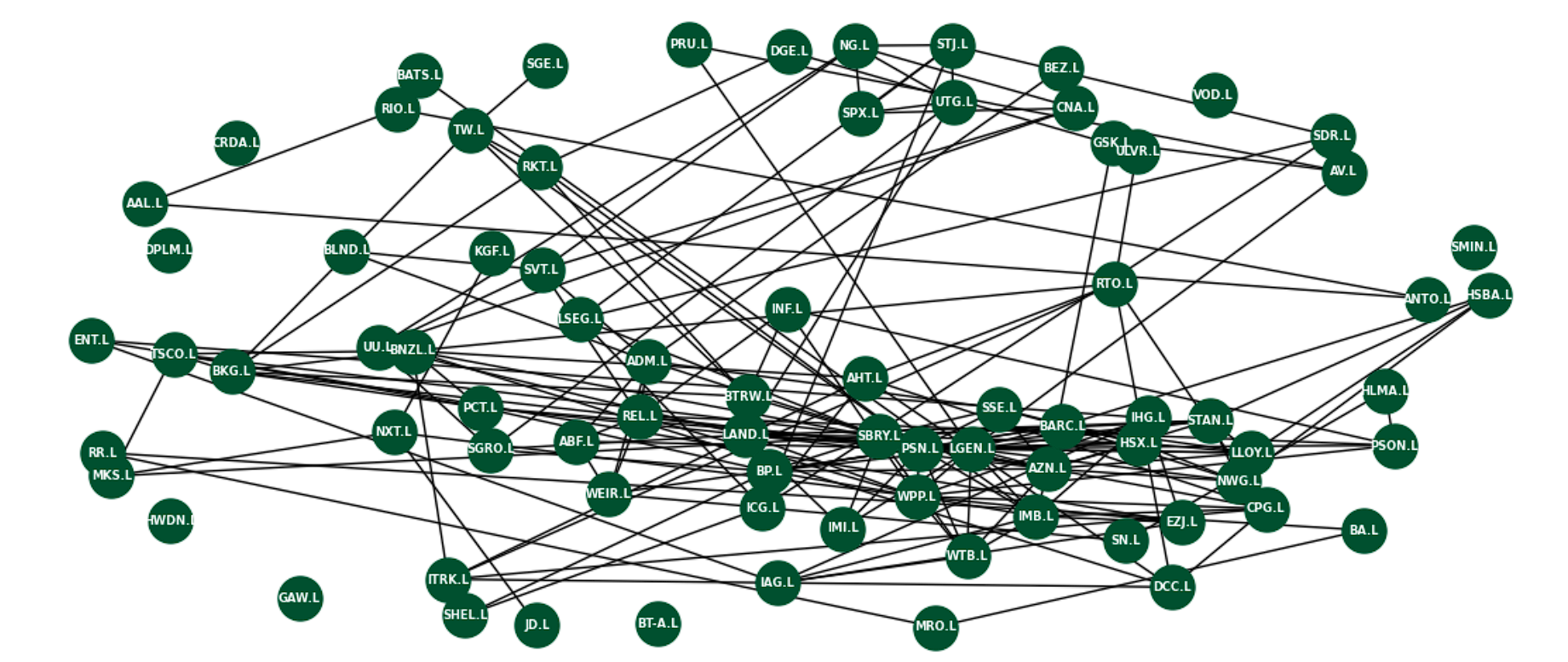}
    \label{fig:placeholder}
\end{figure}
\pagebreak
\noindent Appendix 9: Nodes and edges graphs by sector and Spearman correlation of fundamental ratios
\begin{figure}[H]
    \centering
    \includegraphics[width=0.75\linewidth]{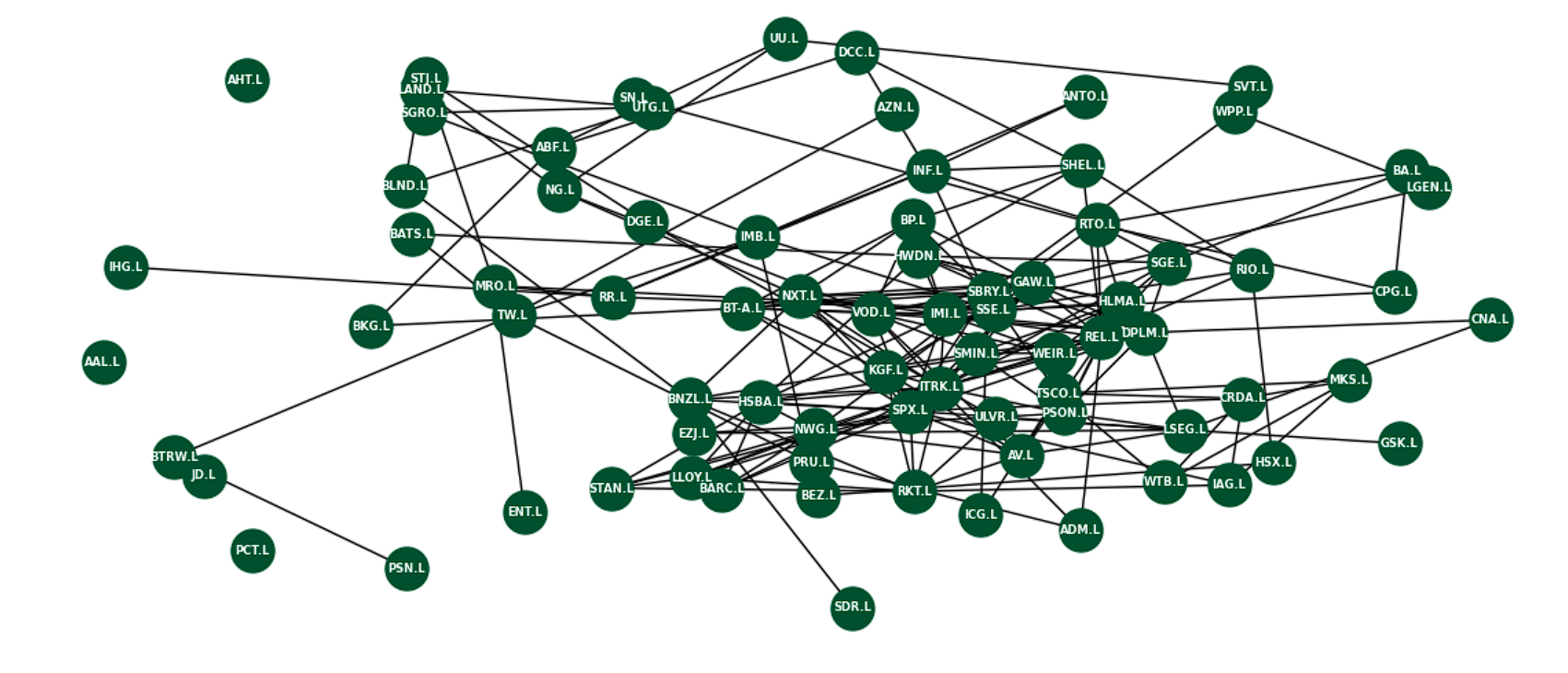}
    \label{fig:placeholder}
\end{figure}

\noindent Appendix 10: Spectral-based GCN

\noindent The previous generation worked as follows: Figure 5 represents how each node aggregates information from neighbouring nodes on a two-layer configuration used in this model. Each layer is the same graph composition superimposed so that each node can aggregate individually with its respective neighbours.
\\

\begin{figure}[H]
    \centering
    \includegraphics[width=0.5\linewidth]{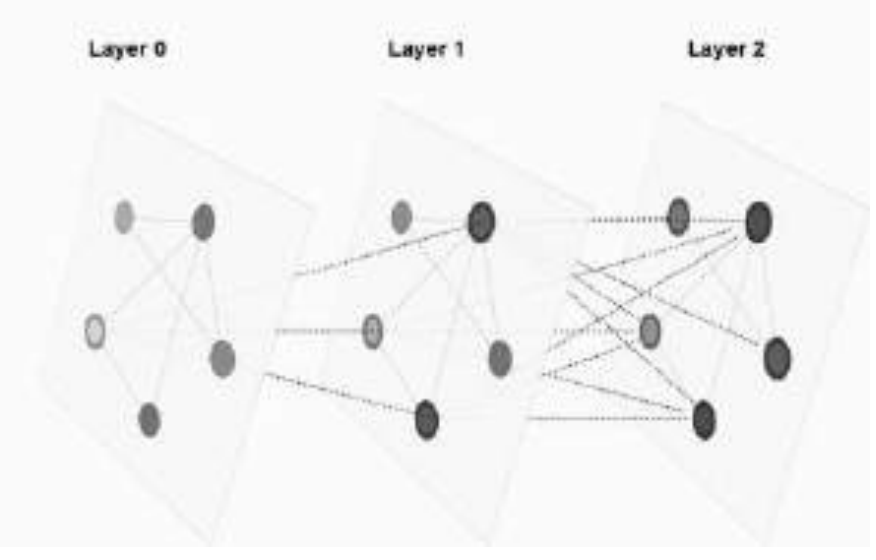}
    \label{fig:placeholder}
\end{figure}

\noindent Appendix 11: MAE, RMSE, MSE, MRE Equations

\begin{align*}
L_{\text{MAE}} &= \frac{1}{n} \sum_{i=1}^n \lvert v_i - \hat{v}_i \rvert \\
L_{\text{RMSE}} &= \frac{1}{n} \sqrt{\sum_{i=1}^n (v_i - \hat{v}_i)^2} \\
L_{\text{MSE}} &= \frac{1}{n} \sum_{i=1}^n (v_i - \hat{v}_i)^2 \\
L_{\text{MRE}} &= \frac{1}{n} \sum_{i=1}^n \frac{\lvert v_i - \hat{v}_i \rvert}{\hat{v}_i}
\end{align*}

\noindent Appendix 12: Individual Plots for predicted vs actual test dataset

\begin{figure}
    \centering
    \includegraphics[width=1\linewidth]{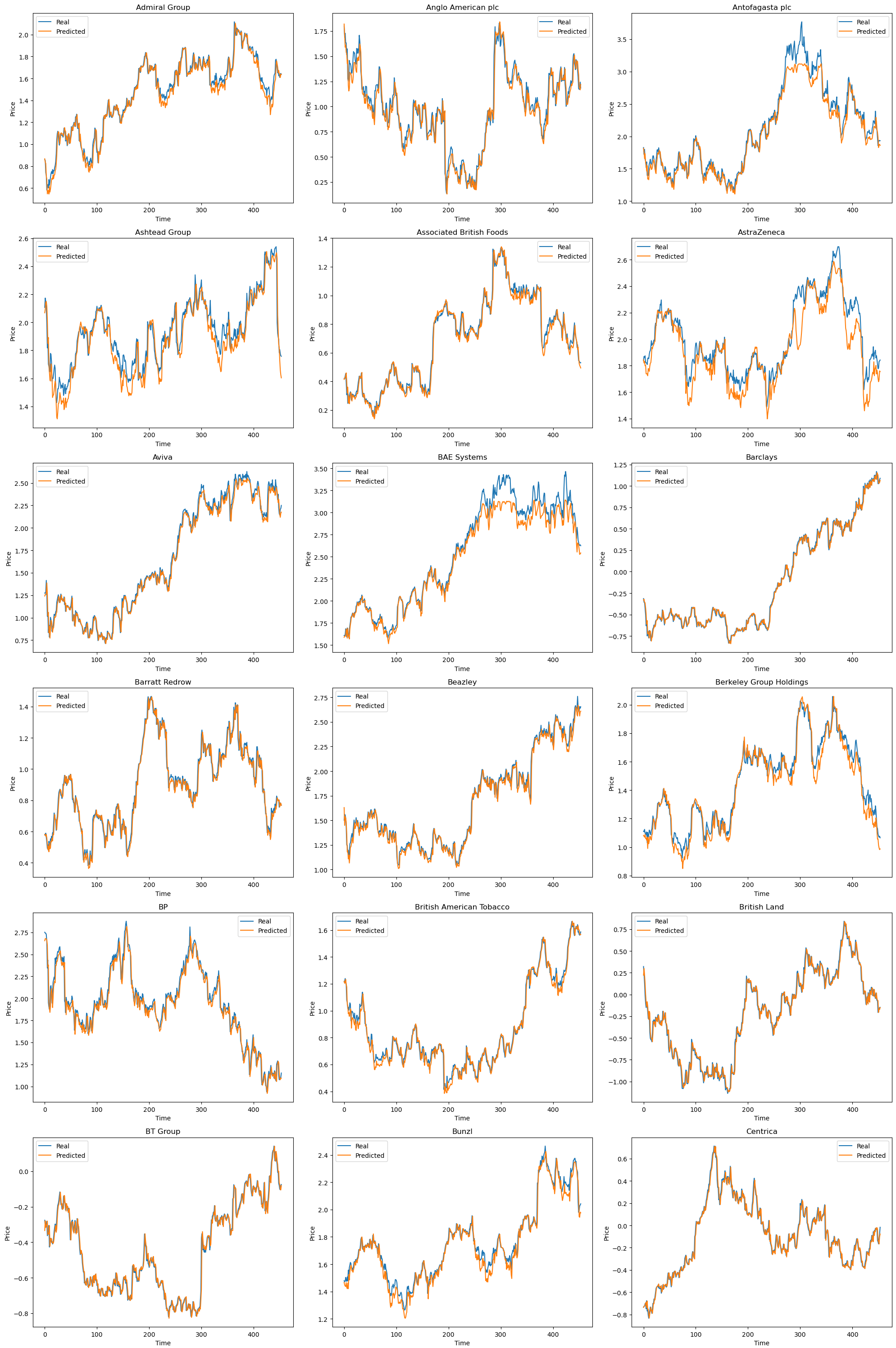}
    \label{fig:placeholder}
\end{figure}

\pagebreak

\end{document}